\documentstyle[prd,aps,preprint,tighten,floats,epsfig]{revtex} 

\begin{document}

\reversemarginpar

\title{On the response of non-linearly coupled, accelerated\\
detectors in odd-dimensional flat spacetimes}
\author{L.~Sriramkumar~\thanks{E-mail:~slakshm@phys.ualberta.ca}}
\address{Theoretical Physics Institute, Department of Physics\\
University of Alberta, Edmonton, Alberta~T6G~2J1, Canada}

\maketitle

\begin{abstract}
In this note, we consider the response of a uniformly accelerated 
monopole detector that is coupled {\it non-linearly}\/ to the~$n$th 
power of a quantum scalar field in $(D+1)$\/-dimensional flat spacetime. 
We show that, when $(D+1)$\/ is even, the response of the detector 
in the Minkowski vacuum is characterized by a Bose-Einstein factor
for all~$n$.\/ 
Whereas, when $(D+1)$\/ is odd, we find that a Fermi-Dirac factor 
appears in the detector response when~$n$\/ is odd, but a Bose-Einstein 
factor arises when~$n$\/ is even.
\end{abstract}

\newpage

It has been a quarter of a century now since it was discovered 
that the response of a uniformly accelerated monopole 
detector that is coupled to a quantized massless scalar field 
is characterized by a Planckian distribution when the field is 
assumed to be in the Minkowski vacuum~\cite{unruh76,dewitt79}.
However, about a decade after the original discovery, it was  
noticed that this result is true only in even-dimensional flat 
spacetimes and it was pointed out that a Fermi-Dirac factor 
(rather than a Bose-Einstein factor) appears in the response of 
the accelerated detector when the dimensionality of spacetime 
is odd (see 
Refs.~\cite{takagi84,takagi85a,stephens85,ooguri86,unruh86,takagi86}; 
for relatively recent discussions, see 
Refs.~\cite{anglin93,terashima99}). 
The detector due to Unruh~\cite{unruh76} and DeWitt~\cite{dewitt79}
is coupled {\it linearly}\/ to the quantum scalar field.
During the last decade or so, motivated by different reasons,
there has been an occasional interest in literature in studying 
the response of detectors that are coupled {\it non-linearly}\/ 
to the quantum 
field~\cite{hinton83,hinton84,paddytp87,suzuki97,sriram99}.
It will be interesting to examine whether the non-linearity 
of the coupling affects the result in odd-dimensional flat 
spacetimes that we mentioned above.

In this note, we shall consider the response of a uniformly 
accelerated monopole detector that is coupled to the $n$th 
power (with $n$ being a positive integer) of a quantum scalar 
field in $(D+1)$-dimensional flat spacetime.
As we shall see, the non-linearity of the coupling affects 
the afore-mentioned result in odd spacetime dimensions in 
an interesting fashion.
We shall show that, when $(D+1)$ is even, a Bose-Einstein 
factor arises in the response of the detector for all $n$, 
whereas, when $(D+1)$ is odd, a Fermi-Dirac factor appears 
in the detector response when~$n$ is odd, but a Bose-Einstein 
factor arises when $n$ is even.

Consider a monopole detector that is moving along a trajectory 
${\tilde x}(\tau)$, where $\tau$ is the proper time in the frame 
of the detector.
Let the detector interact with a real scalar field~$\Phi$ through 
the non-linear interaction Lagrangian~\cite{suzuki97}
\begin{equation}
{\cal L}_{\rm int}
= c\, m(\tau)\; \Phi^n\left[{\tilde x}(\tau)\right],
\label{eqn:nlint}
\end{equation}
where $c$ is a small coupling constant, $m(\tau)$ is 
the detector's monopole moment and~$n$ is a positive
integer that denotes the index of non-linearity of 
the coupling.
Let us now assume that the quantum field ${\hat \Phi}$ is 
initially in the vacuum state $\left\vert 0 \right\rangle$ 
and the detector is in its ground state~$\left\vert E_0 
\right\rangle$ 
corresponding to an energy eigen value~$E_0$.
Then, up to the first order in perturbation theory, the 
amplitude of transition of the non-linearly coupled detector 
to an excited state~$\left\vert E \right\rangle$, corresponding 
to an energy eigen value~$E\, \left(>E_0\right)$, is described 
by the integral (see, for e.g., Ref.~\cite{bd82})
\begin{equation}
{\cal A}_{n}({\cal E}) = {\cal M}
\int\limits_{-\infty}^{\infty} d\tau\, e^{i {\cal E}\tau}\, 
\left\langle\Psi\right\vert\, {\hat \Phi}^n[{\tilde x}(\tau)]\,
\left\vert 0\right\rangle,
\label{eqn:detamp}
\end{equation}
where ${\cal M}\equiv \left(ic\, \left\langle E \right\vert 
{\hat m}(0)\left\vert E_{0} \right\rangle\right)$, ${\cal E}
=\left(E-E_0\right)>0$ and $\left\vert \Psi \right\rangle$ is 
the state of the quantum scalar field after its interaction 
with the detector.
(Since the quantity ${\cal M}$ depends only on the internal 
structure of the detector and does not depend on its motion, 
we shall drop this quantity hereafter.) 

The transition amplitude~${\cal A}_{n}({\cal E})$ above 
involves products of the quantum field~${\hat \Phi}$ at the 
{\it same}\/ spacetime point and, hence, we will encounter 
divergences when evaluating this transition amplitude.
In order to avoid the divergences, we shall normal order 
the operators in the matrix element in the transition 
amplitude~${\cal A}_{n}({\cal E})$ with respect to the 
Minkowski vacuum~\cite{suzuki97}.
That is, we shall assume that the transition 
amplitude~(\ref{eqn:detamp}) above is instead 
given by the expression
\begin{equation}
{\tilde {\cal A}}_{n}({\cal E}) 
=\int\limits_{-\infty}^{\infty} d\tau\, e^{i {\cal E}\tau}\, 
\left\langle\Psi\right\vert
:{\hat \Phi}^n[{\tilde x}(\tau)]:\left\vert 0\right\rangle,
\label{eqn:nodetamp}
\end{equation}
where the colons denote normal ordering with respect to the
Minkowski vacuum.
Then, the transition probability of the detector to all 
possible final states $\left\vert \Psi\right\rangle$ of 
the quantum field is given by
\begin{equation}
{\cal P}_{n}({\cal E}) 
=\sum_{\vert\Psi\rangle}{\vert 
{\tilde {\cal A}}_{n}({\cal E})\vert}^2
= \int\limits_{-\infty}^\infty d\tau\, 
\int\limits_{-\infty}^\infty d\tau'\, 
e^{-i{\cal E}(\tau-\tau')}\, 
G^{(n)}\left[{\tilde x}(\tau), {\tilde x}(\tau')\right],
\label{eqn:detprob}
\end{equation}
where $G^{(n)}\left[{\tilde x}(\tau), {\tilde x}(\tau')\right]$ 
is the $(2n)$-point function defined as
\begin{equation}
G^{(n)}\left[{\tilde x}(\tau), {\tilde x}(\tau')\right]
=\left\langle 0 \right\vert 
:{\hat \Phi}^n\left[{\tilde x}(\tau)\right]:\,
:{\hat \Phi}^n\left[{\tilde x}(\tau')\right]:
\left\vert 0 \right\rangle.\label{eqn:2nptfn}
\end{equation}
In cases wherein the  $(2n)$-point function 
$G^{(n)}\left(\tau, \tau'\right) \left(\equiv G^{(n)}
\left[{\tilde x}(\tau), {\tilde x}(\tau')\right]\right)$ is 
invariant under time translations in frame of the detector, 
we can define a transition probability rate for the detector 
as follows:
\begin{equation}
{\cal R}_{n}({\cal E}) 
= \int\limits_{-\infty}^\infty d{(\tau - \tau')}\;
e^{-i{\cal E}(\tau - \tau')}\; G^{(n)}(\tau - \tau').
\label{eqn:detrate}
\end{equation} 

Let us now assume that the quantum scalar field~${\hat \Phi}$ 
is in the Minkowski vacuum.
In such a case, the $(2n)$-point function $G^{(n)}\left({\tilde x}, 
{\tilde x'}\right)$ simplifies to
\begin{equation}
G^{(n)}_{\rm M}\left({\tilde x}, {\tilde x'}\right)
= \left(n!\right)\, \left[G^{+}_{\rm M}\left({\tilde x}, 
{\tilde x'}\right)\right]^n,\label{eqn:2nptfnmv}
\end{equation}
where $G^{+}_{\rm M}\left({\tilde x}, {\tilde x'}\right)$ 
denotes the Wightman function in the Minkowski vacuum\footnote{It 
ought to be noted here that we would have arrived at the 
expression~(\ref{eqn:2nptfnmv}) for the $(2n)$-point function in 
the Minkowski vacuum even if we had started with the transition 
amplitude~(\ref{eqn:detamp}) (instead of the normal ordered 
amplitude~(\ref{eqn:nodetamp})), rewritten the resulting $(2n)$-point 
function in the transition probability in terms of the two-point 
functions using Wick's theorem and then replaced the divergent 
terms that arise (i.e. those two-point functions with coincident 
points) with the corresponding regularized expressions (for a 
discussion on this point, also see Ref.~\cite{suzuki97}).}.
In $(D+1)$ spacetime dimensions (and for $(D+1)\ge 3$), the 
Wightman function for a massless scalar field in the Minkowski 
vacuum is given by (see, for instance, 
Refs.~\cite{takagi85a,unruh86,takagi86})
\begin{equation}
G^{+}_{\rm M}({\tilde x},{\tilde x'})
= {\cal C}_{D}\; \left[(-1)\;
\left((t-t'-i\epsilon)^2
-\vert {\bf x}-{\bf x}'\vert^2\right)\right]^{-(D-1)/2},
\label{eqn:wgfnmvine}
\end{equation}
where $\epsilon\to 0^{+}$, $\left[t, {\bf x}\equiv\left(x^{1},
x^{2},\ldots,x^{D}\right)\right]$ are the Minkowski coordinates 
and the quantity~${\cal C}_{D}$ is given by  
\begin{equation}
{\cal C}_{D}= \left(4\pi^{(D+1)/2}\right)^{-1}\;
\Gamma\left[(D-1)/2\right]
\label{eqn:cd}
\end{equation}
with $\Gamma\left[(D-1)/2\right]$ denoting the Gamma function.

Now, consider a detector accelerating uniformly along the 
$x^{1}$~direction with a proper acceleration~$g$. 
The trajectory of such a detector is given by (see, for e.g.,
Ref.~\cite{bd82})
\begin{equation}
t(\tau)=g^{-1}\, {\rm sinh}(g\tau)\;\;,\;\; 
x^{1}(\tau)=g^{-1}\, {\rm cosh}(g\tau)\;\;,\;\;
x^{2}=x^{3}=\ldots=x^{D}=0,
\label{eqn:acctraj}
\end{equation}
where~$\tau$ is the proper time in the frame of the detector.
On substituting this trajectory in the Minkowski Wightman 
function~(\ref{eqn:wgfnmvine}), we obtain that (see, for e.g, 
Refs.~\cite{takagi85a,unruh86,takagi86})
\begin{equation}
G_{\rm M}^{+}(\tau, \tau')
=\left[{\cal C}_{D}\; (g/2i)^{(D-1)}\right]\;
\biggl({\rm sinh}\left[\left(g(\tau-\tau')/2\right)
-i\epsilon\right]\biggl)^{-(D-1)}.
\label{eqn:wgfnmvrind}
\end{equation}
Therefore, along the trajectory of the uniformly 
accelerated detector, the $(2n)$-point function in 
the Minkowski vacuum~(\ref{eqn:2nptfnmv}) is given by
\begin{equation}
G^{(n)}_{\rm M}\left(\tau, \tau'\right)
=(n!)\;\left[{\cal C}_{D}^{n}\; (g/2i)^{\alpha}\right]\; 
\biggl({\rm sinh}\left[\left(g(\tau-\tau')/2\right)
-i\epsilon\right]\biggl)^{-\alpha},
\label{eqn:2nptfnmvrind}
\end{equation}
where $\alpha=\left[(D-1)n\right]$.

On substituting the $(2n)$-point 
function~(\ref{eqn:2nptfnmvrind}) in the 
expression~(\ref{eqn:detrate}) and carrying 
out the resulting integral, we find that the transition 
probability rate of the uniformly accelerated, non-linearly
coupled detector can be written as (cf.~Ref.~\cite{gr80}, p.~305, 
Eq.~3.314; p.~950, Eq.~8.384.1; p.~937, Eqs.~8.331, 8.332.1 
and 8.332.2)
\begin{equation}
{\cal R}_{n}({\cal E}) 
={\cal B}(n,D)\;\;
\left\{
\begin{array}{l}
\left(g^{\alpha}/{\cal E}\right)\;
{\underbrace{\left[\exp(2\pi{\cal E}/g)-1\right]^{-1}}}\;
\prod\limits_{l=0}^{(\alpha-2)/2}
\left[l^2+({\cal E}/g)^2\right]\\
\qquad\quad\,\mbox{Bose-Einstein factor}\qquad
\qquad\qquad\quad\;\mbox{when $\alpha$ is even}\\
g^{(\alpha-1)}\;\;
{\underbrace{\left[\exp(2\pi{\cal E}/g)+1\right]^{-1}}}\;
\prod\limits_{l=0}^{(\alpha-3)/2}
\left[((2l+1)/2)^2+({\cal E}/g)^2\right]\\
\qquad\quad\;\mbox{Fermi-Dirac factor}\qquad
\qquad\qquad\qquad\mbox{when $\alpha$ is odd,}
\end{array}\right.
\label{eqn:fnlexp}
\end{equation}
where the quantity ${\cal B}(n,D)$ is given by
\begin{equation}
{\cal B}(n,D)=(2\pi)\,(n!)\; 
\left[{\cal C}_{D}^{n}/\Gamma(\alpha)\right].
\end{equation}
When $(D+1)$ is even, $\alpha$ is even for all $n$ and, hence, 
a Bose-Einstein factor will always arise in the response of 
the uniformly accelerated detector in an even-dimensional flat 
spacetime.
Whereas, when $(D+1)$ is odd, evidently, $\alpha$ will be odd 
or even depending on whether $n$ is odd or even.
Therefore, in an odd-dimensional flat spacetime, a Fermi-Dirac 
factor will arise in the detector response when~$n$ is odd (as
in the case of the Unruh-DeWitt detector), but a Bose-Einstein 
factor will appear when~$n$ is even!

\begin{figure}[!htb]
\vskip 20 pt
$\qquad\qquad\qquad\!\!{\bar {\cal R}}_{n}({\cal E})$
\begin{center}
\vskip -20 pt
\epsfbox{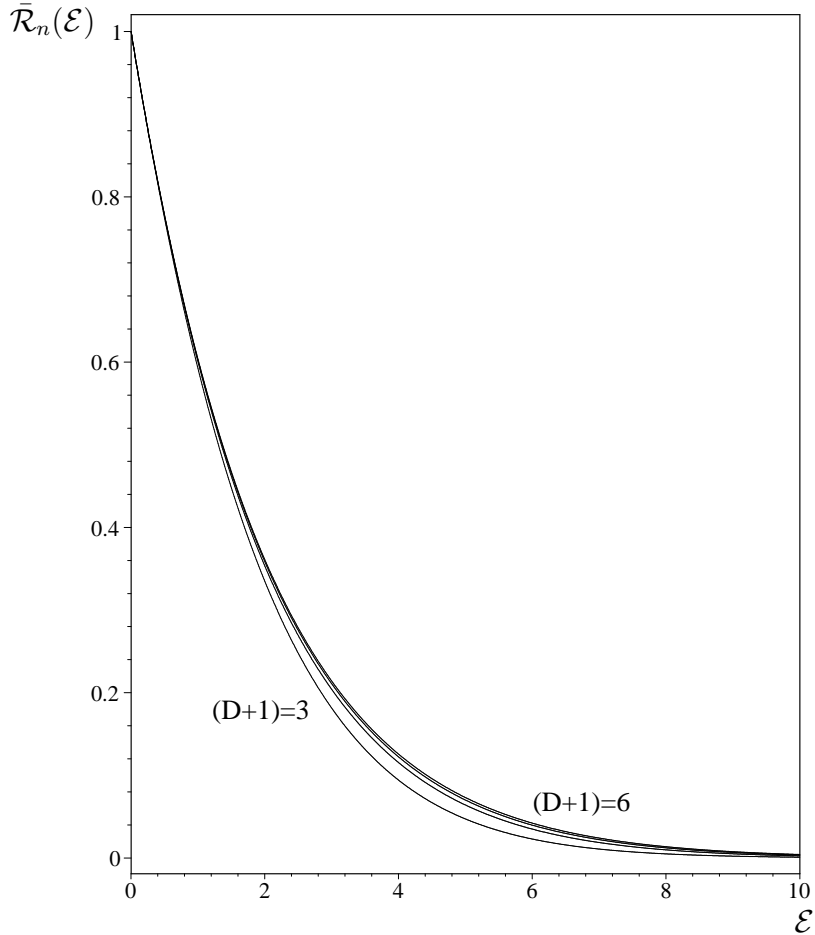}
\hskip -10 pt ${\cal E}$
\end{center}
\vskip 10 pt
\centerline{\caption{${\bar {\cal R}}_{n}({\cal E})\,\, 
{\rm vs.}\,\,{\cal E}$ for $n=3$ and $(D+1)=3,4,5,6$.}
\label{fig:n3}}
\end{figure}

\begin{figure}[!htb]
\vskip 20 pt
$\qquad\qquad\qquad\!\!{\bar {\cal R}}_{n}({\cal E})$
\begin{center}
\vskip -20 pt
\epsfbox{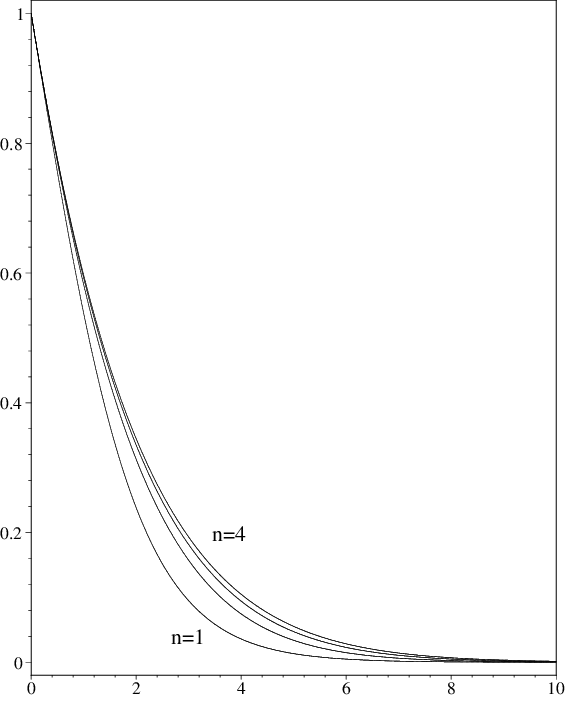}
\hskip -10 pt ${\cal E}$
\end{center}
\vskip 10 pt
\centerline{\caption{${\bar {\cal R}}_{n}({\cal E})\,\, 
{\rm vs.}\,\, {\cal E}$ for $(D+1)=3$ and $n=1,2,3,4$.}
\label{fig:D2}}
\end{figure}

Three points need to be emphasized regarding this curious result.
Firstly, the temperature associated with the Bose-Einstein and 
the Fermi-Dirac factors that appear in the response of the
non-linearly coupled detector is the standard Unruh temperature, 
viz.~$(g/2\pi)$. 
Secondly, the response of the detector is characterized 
{\it completely}\/ by either a Bose-Einstein or a Fermi-Dirac 
distribution {\it only}\/ in situations wherein $\alpha<3$.
For cases such that $\alpha\ge 3$, the detector response contains,
in addition to a Bose-Einstein or a Fermi-Dirac factor, a term 
which is polynomial in $({\cal E}/g)$.
Thirdly, in Figs.~\ref{fig:n3} and~\ref{fig:D2}, 
following Unruh~\cite{unruh86}, we have plotted the transition 
probability rate of the detector (in fact, the quantity 
${\bar {\cal R}}_{n}({\cal E})=\left[{\cal R}_{n}({\cal E})/
{\cal R}_{n}(0)\right]$) for a few different values of~$D$ for 
the case $n=3$ and for a few different values of~$n$ for the 
case wherein $(D+1)=3$, respectively.
(In plotting these figures, we have set $g=(2\pi)$.)
It is interesting to note from these two figures that, though the 
characteristic response of the detector alternates between the 
Bose-Einstein and the Fermi-Dirac factors as we go from one $D$ 
to another for odd $n$ (or from one $n$ to another when $(D+1)$ 
is odd), the complete spectra themselves exhibit a smooth  
dependence both on the index of non-linearity of the coupling and 
the dimension of spacetime.

The appearance of a Fermi-Dirac factor (instead of the expected 
Planckian distribution) has been shown to occur in the response 
of a comoving Unruh-DeWitt detector (that is coupled to a massless 
scalar field) in odd-dimensional de Sitter spacetimes (see, for 
e.g., Refs.~\cite{ooguri86,takagi85b}; for a recent discussion, 
see Ref.~\cite{mty01}) and also in the case of a detector stationed 
at a constant radius in the spacetime of the $(2+1)$~dimensional 
BTZ black hole~\cite{lo94}. 
Moreover, it has recently been pointed out that, not only comoving,
but even accelerated Unruh-DeWitt detectors in de Sitter spacetime 
(and, also detectors with proper acceleration beyond a certain 
critical value in anti-de Sitter spacetime) exhibit a thermal 
response~\cite{dl97,jacobson98,dl98,dl99}. 
It will be interesting to investigate as to how the non-linearity
of the detector's coupling would affect the response of a static
detector around the BTZ black hole and also the response of comoving 
as well as accelerated detectors in de Sitter and anti-de Sitter 
spacetimes in different dimensions.
Furthermore, it has been shown that a similar ``inversion of 
statistics" occurs in the response of a monopole detector that 
is coupled linearly to the scalar density of a massless Dirac 
field in odd-dimensional flat and de Sitter 
spacetimes~\cite{takagi86,terashima99,takagi85b,mty01}
and also around the BTZ black hole~\cite{hsy95}, i.e. the response 
of the detector exhibits a Bose-Einstein factor instead of the 
Fermi-Dirac factor expected in such situations.
It will be worthwhile to examine as to how detectors coupled
non-linearly to the scalar density of spinor fields respond 
in odd-dimensional spacetimes. 
We plan to address these issues in some detail in a forthcoming 
publication~\cite{sriramwp}.

\acknowledgements 
We would like to thank Don Page, Valeri Frolov and Jonathan 
Oppenheim for discussions, William Unruh for correspondence 
and discussions, Andrei Zelnikov for comments on the manuscript
and Supratim Sengupta for help with Maple and RevTeX.
This work was supported by the Natural Sciences and Engineering 
Research Council of Canada. 


\references
\bibitem{unruh76}
W.~G.~Unruh, Phys.\ Rev.\ D\ {\bf 14}, 870 (1976).
\bibitem{dewitt79}
B.~S.~DeWitt, {\it Quantum gravity:~The new synthesis}, 
in {\sl General Relativity:~An Einstein Centenary Survey}, 
Eds. S.~W.~Hawking and W.~Israel (Cambridge University 
Press, Cambridge, England, 1979).
\bibitem{takagi84}
S.~Takagi, Prog.\ Theor.\ Phys.\ {\bf 72}, 505 (1984).
\bibitem{takagi85a}
S.~Takagi, Prog.\ Theor.\ Phys.\ {\bf 74}, 142 (1985).
\bibitem{stephens85}
C.~R.~Stephens, University of Maryland Report, 1985 
(Unpublished).
\bibitem{ooguri86}
H.~Ooguri, Phys.\ Rev.\ D {\bf 33}, 3573 (1986).
\bibitem{unruh86}
W.~G.~Unruh, Phys.\ Rev.\ D\ {\bf 34}, 1222 (1986).
\bibitem{takagi86}
S.~Takagi, Prog.\ Theor.\ Phys.\ Suppl.\ {\bf 88}, 1 (1986).
\bibitem{anglin93}
J.~R.~Anglin, Phys.\ Rev.\ D {\bf 47}, 4525 (1993).
\bibitem{terashima99}
H.~Terashima, Phys.\ Rev.\ D {\bf 60}, 084001 (1999).
\bibitem{hinton83}
K.~J.~Hinton, J.\ Phys.\ A:\ Math.\ Gen.\ {\bf 16}, 
1937 (1983).
\bibitem{hinton84}
K.~J.~Hinton, Class.\ Quantum Grav.\ {\bf 1}, 27 (1984).
\bibitem{paddytp87}
T.~Padmanabhan and T.~P.~Singh, Class.\ Quantum Grav.\ 
{\bf 4}, 1397 (1987).
\bibitem{suzuki97}
N.~Suzuki, Class.\ Quantum Grav.\ {\bf 14}, 3149 (1997).
\bibitem{sriram99}
L.~Sriramkumar, Mod.\ Phys.\ Lett.\ A, {\bf 14}, 1869 (1999). 
\bibitem{bd82}
N.~D.~Birrell and P.~C.~W.~Davies, {\sl Quantum Fields in
Curved Space} (Cambridge University Press, Cambridge, England, 
1982), Sec.~3.3.
\bibitem{gr80}
I.~S.~Gradshteyn and I.~M.~Ryzhik, {\sl Table of Integrals,
Series and Products} (Academic Press, New York, 1980). 
\bibitem{takagi85b}
S.~Takagi, Prog.\ Theor.\ Phys.\ {\bf 74}, 501 (1985).
\bibitem{mty01}
T.~Murata, K.~Tsunoda and K.~Yamamoto, gr-qc/0102080.
\bibitem{lo94}
G.~Lifschytz and M.~Ortiz,  Phys.\ Rev.\ D\ {\bf 49}, 1929 (1994).
\bibitem{dl97}
S.~Deser and O.~Levin, Class.\ Quantum Grav.\ {\bf 14},
L163 (1997).
\bibitem{jacobson98}
T.~Jacobson, Class.\ Quantum Grav.\ {\bf 15}, 251 (1998).
\bibitem{dl98}
S.~Deser and O.~Levin,  Class.\ Quantum Grav.\ {\bf 15},
L85 (1998).
\bibitem{dl99}
S.~Deser and O.~Levin, Phys.\ Rev.\ D\ {\bf 59}, 064004 (1999).
\bibitem{hsy95}
S.~Hyun, Y.-S.~Song and J.~H.~Yee,  Phys.\ Rev.\ D\ {\bf 51}, 
1787 (1995).
\bibitem{sriramwp}
L.~Sriramkumar, {\sl Notes on non-linearly coupled
detectors},\/ Work in progress.
\end{document}